\documentstyle[aps]{revtex}

\begin{document}
\title{Boundary Effects in the Magnetic Catalysis of Chiral Symmetry Breaking}
\author{E. J. Ferrer$^{1}$, V. P. Gusynin$^{2}$ and V. de la Incera$^{1}$}
\address{$^{1}${\it Department of Physics, State University of New York, Fredonia, }\\
{\it NY 14063, USA}\\
$^{2}${\it Bogolyubov Institute for Theoretical Physics, 252143 }Kiev, {\it %
Ukraine }}
\date{SUNY-FRE-99-02}
\maketitle

\begin{abstract}
The catalysis of chiral symmetry breaking by an applied constant magnetic
field and in the presence of boundaries along third axis is investigated in
the four-dimensional Nambu-Jona-Lasinio model. It is shown that in case of
periodic boundary conditions for fermions the magnetic field breaks the
chiral symmetry, generating a dynamical mass even at the weakest attractive
interaction between fermions. For antiperiodic boundary conditions the
effect of the finite third dimension is to counteract the chiral symmetry
breaking.
\end{abstract}

\vskip5mm It is a well established result that the global properties of the
space-time, even if it is locally flat, can give rise to new physics. The
seminal discovery in this direction is the so called Casimir effect, that
is, the existence of an attractive force between neutral parallel perfectly
conducting plates\cite{Casimir}. In this phenomenon the attractive force
between the plates is mediated by the zero-point fluctuations of the
electromagnetic field in vacuum. Thus, these Casimir forces are interpreted
as a macroscopic manifestation of the vacuum structure of the quantized
fields in the presence of domains restricted by boundaries or nontrivial
topologies.

From the underlying physical mechanism, it is clear that the Casimir effect
has a broad range of applications. Today, we can find research activity on
this field in many different areas as statistical physics, condensed matter,
elementary particles, cosmology, etc.\cite{Mostepanenko}.

On the other hand, it has been recently found that a magnetic field can
catalyze the dynamical chiral symmetry breaking in different quantum field
systems\cite{CSB1,GMSPL,CSB2} (see, also \cite{others}). This is a universal
phenomenon that can be understood as the generation, through the infrared
dynamics of the fermion pairing in a magnetic field, of a fermion dynamical
mass at the weakest attractive interaction between fermions. The essence of
this phenomenon lies in the dimensional reduction of the electron dynamics
when their energy is much less than the Landau gap $\sqrt{eB}$ ($B$ is the
magnitude of the magnetic field)\cite{CSB1,GMSPL,CSB2}. In this case, the
electrons are confined to the lowest Landau level, therefore having a
(D-2)-dimensional dynamics. The lowest Landau level plays in this case a
role similar to that of the Fermi surface in BCS superconductivity.

As it has been pointed out by several authors, the magnetic catalysis of
dynamical chiral symmetry breaking can have important applications in
condensed matter physics\cite{CM}$,$ quantum chromodynamics\cite{smilga} and
cosmology\cite{CSB2,Cosmology}.

Our main goal in this paper is to combine these two effects, that is, to
investigate the chiral symmetry breaking in the presence of an external
constant magnetic field for a fermion system in a space-time which is
locally flat but which has a nontrivial topology represented by the domain $%
R^{3}\times S^{1}$ (i.e. a Minkowskian space with one of the spatial
dimensions compactified in a circle $S^{1}$ of finite length $a$).

The fact that the energy of the vacuum state of quantized matter fields in $%
R^{3}\times S^{1}$ is nonzero has been corroborated by several authors in
different physical models\cite{0-Energy,Ford}. Recently, the influence of a
chemical potential \cite{Klimenko}, as well as of an external magnetic field 
\cite{Farina}, on the Casimir energy density of charged particles (bosonic
and fermionic) has been also investigated.

We start considering the Lagrangian density of free fermions in the presence
of a constant external magnetic field 
\begin{equation}
{\cal L}=\frac{1}{2}\left[ \overline{\psi },\left( i\gamma ^{\mu }D_{\mu
}-m\right) \psi \right] ,\qquad \mu =0,1,2,3,  \label{1}
\end{equation}
where the covariant derivative $D_{\mu }=\partial _{\mu }-ieA_{\mu }^{ext}$
depends on the external potential $A_{\mu }^{ext},$ chosen in the Landau
gauge 
\begin{equation}
A_{\mu }^{ext}=-\delta _{\mu 1}Bx_{2}.  \label{3}
\end{equation}
This potential corresponds to a constant magnetic field pointing in the $%
{\cal OZ}$ positive direction.

The nontrivial topology of the compactified space-time domain $R^{3}\times
S^{1}$ is transferred into periodic boundary conditions ({\it PBC}) for
untwisted fermions or antiperiodic boundary conditions ({\it APBC}) for
twisted fermions\cite{Ford}\footnote{%
For a discussion of the role of the boundary conditions of the fermion
fields on the dynamical symmetry breaking in flat space see, for example,
ref.\cite{Ishikawa}. Curvature effects (as well as curvature in combination
with an external magnetic field) on the symmetry breaking have been also
studied thoroughly ( for an excellent review and extended list of references
see \cite{Muta}).} 
\begin{equation}
\psi \left( t,x,y,z+a\right) =\pm \psi \left( t,x,y,z\right) .  \label{4}
\end{equation}
In (\ref{4}) we have taken the compactified dimension with length $a$ in the 
${\cal OZ}$-direction.

The fermion energy eigenvalues corresponding to the Lagrangian density (\ref
{1}) with {\it PBC } and {\it APBC } are respectively given by 
\begin{equation}
E_{n,l}^{(P)}=\pm \sqrt{m^{2}+2\left| eB\right| l+\frac{4\pi ^{2}}{a^{2}}%
n^{2}}\qquad l=0,1,2,...,\qquad n=0,\pm 1,\pm 2,...  \label{6}
\end{equation}
\begin{equation}
E_{n,l}^{(AP)}=\pm \sqrt{m^{2}+2\left| eB\right| l+\frac{4\pi ^{2}}{a^{2}}%
\left( n+\frac{1}{2}\right) ^{2}}\qquad l=0,1,2,...,\qquad n=0,\pm 1,\pm
2,....  \label{7}
\end{equation}
In Eqs. (\ref{6}) and (\ref{7}) $l$ represents the Landau level and $n$ the
discrete components of $P_{z}$, the momentum component in the direction of
the compactified spatial coordinate.

From Eq. (\ref{6}) we see that for the {\it PBC} case there is no energy gap
between the vacuum and the lowest Landau level ($l=0$) at zero momentum ($%
n=0 $) in the $m\rightarrow 0$ limit. This is the same behavior found in
this system in (2+1)- and (3+1)-dimensions, when it is considered in a
topologically trivial space-time ($a\rightarrow \infty $) \cite
{CSB1,GMSPL,CSB2}. As in those cases, it should be expected here that the
magnetic field catalyzes the dynamical breaking of chiral symmetry. The
absence of an energy gap between the vacuum and the lowest Landau level at
zero momentum makes possible the vacuum condensation of electron pairs which
are interacting in the infrared band of the fermion spectrum.

For the {\it APBC} case (Eq. (\ref{7})) the situation is different. Even at $%
m\rightarrow 0$, an energy gap $\left( \Delta E=\frac{\pi }{a}\right) $,
which depends inversely on the compactified dimension length $a$, exists
between the vacuum and the lowest Landau level ($l=0$) at zero momentum ($%
n=0 $).

In each case, to determine whether a chiral condensate $\left\langle 0\left| 
\overline{\psi }\psi \right| 0\right\rangle $ catalyzed by the applied
external field exists, we start from the fermion effective action in the
presence of a constant magnetic field which in the Schwinger proper time
formalism \cite{Schwinger} is given by 
\begin{equation}
W=\frac{i}{2}\int\limits_{1/\Lambda^2}^{\infty }\frac{ds}{s}{\rm Tr}e^{-is%
{\cal H} },  \label{5a}
\end{equation}
where $1/\Lambda^2$ is a cutoff in the proper time $s$, ${\rm Tr}$ means the
trace over space-time and internal degrees of freedom, and the proper-time
Hamiltonian density is 
\begin{equation}
{\cal H}=\left( P_{\mu }-eA_{\mu }^{ext}\right) ^{2}-\frac{e}{2}\sigma _{\mu
\nu }F^{\mu \nu }+m^{2}  \label{5}
\end{equation}
with $P_{\mu }=-i\partial _{\mu }$ denoting the electron four-momentum and $%
\sigma _{\mu \nu }=\frac{i}{2}\left[ \gamma _{\mu },\gamma _{\nu }\right] $.

For the {\it APBC} case it was found in the second paper of Ref. \cite
{Farina} that the fermion effective action is given by 
\begin{equation}
W_{AP}=\frac{T{\cal A}a}{8\pi ^{2}}\int\limits_{1/\Lambda ^{2}}^{\infty }%
\frac{ds}{s^{3}}e^{-ism^{2}}\left[ 1+2\sum\limits_{n=1}^{\infty }\left(
-1\right) ^{n}e^{i\left( an\right) ^{2}/4s}\right] \left[ 1+iseB{\it L}%
\left( iseB\right) \right] .  \label{8}
\end{equation}
In Eq. (\ref{8}) the sum over the Landau levels was carried out, $V=T{\cal A}%
a$ is the four-dimensional volume of the $R^{3}\times S^{1}$ domain and $%
{\it L}\left( \xi \right) =\coth \xi -\xi ^{-1}$ is the Langevin function.

In case of periodic boundary conditions the only difference in the effective
action $W_{P}$ as compared to $W_{AP}$ is the absence of the phase factor $%
(-1)^{n}$ appearing in the sum of Eq.(\ref{8}). Thus the effective action
for the {\it PBC} case is 
\begin{equation}
W_{P}=\frac{T{\cal A}a}{8\pi ^{2}}\int\limits_{1/\Lambda^2}^{\infty }\frac{ds%
}{s^{3}}e^{-ism^{2}}\left[ 1+2\sum\limits_{n=1}^{\infty }e^{i\left(
an\right) ^{2}/4s}\right] \left[ 1+iseB{\it L}\left( iseB\right) \right] .
\label{9}
\end{equation}
The chiral condensate is found through the equation 
\begin{equation}
\frac{d{\cal W}}{dm}\Big|_{m=0} =-\left\langle 0\left| \overline{\psi }\psi
\right| 0\right\rangle.  \label{10}
\end{equation}
where ${\cal W}=\frac{W}{T{\cal A}a}$. From Eqs. (\ref{8}) and (\ref{9}) we
have respectively 
\begin{equation}
\frac{d{\cal W}_{AP}}{dm}=-\frac{m}{4\pi ^{2}}\int\limits_{1/\Lambda^2}^{%
\infty } \frac{ds}{s^{2}}e^{-sm^{2}}\theta _{4}\left( 0\left| \frac{ia^{2}}{%
4\pi s} \right. \right) \left[ 1+seB{\it L}\left( seB\right) \right] ,
\label{11}
\end{equation}

\begin{equation}
\frac{d{\cal W}_{P}}{dm}=-\frac{m}{4\pi ^{2}}\int\limits_{1/\Lambda^2}^{%
\infty }\frac{ds}{s^{2}}e^{-sm^{2}}\theta _{3}\left( 0\left| \frac{ia^{2}}{%
4\pi s}\right. \right) \left[ 1+seB{\it L}\left( seB\right) \right] ,
\label{12}
\end{equation}
where we have introduced the Jacobi $\theta $-functions\cite{Table} to
represent the series appearing in Eqs. (\ref{8}) and (\ref{9}), and made a
Wick rotation to Euclidean space $\left( s\rightarrow -is\right) $.

Using the Jacobi imaginary transformations 
\begin{equation}
\theta _{4}\left( 0\left| \tau \right. \right) =\sqrt{\frac{i}{\tau }}
\theta_{2}\left( 0\left| -\frac{1}{\tau }\right. \right),\quad \theta
_{3}\left( 0\left| \tau \right. \right) =\sqrt{\frac{i}{\tau }}
\theta_{3}\left( 0\left|- \frac{1}{\tau }\right. \right)  \label{13}
\end{equation}
with $\tau =\frac{ia^{2}}{4\pi s}$, we can write Eqs. (\ref{11}) and (\ref
{12}) as 
\begin{eqnarray}
\frac{d{\cal W}_{AP}}{dm} &=&-\frac{eB}{\pi ^{3/2}a}\int\limits_{m^{2}/
\Lambda ^{2}}^{\infty }\frac{ds}{s^{1/2}}e^{-s}\sum\limits_{n=1}^{\infty
}e^{-s\frac{4\pi ^{2}\left( n+1/2\right) ^{2}}{\left( am\right) ^{2}}}\coth
\left( s\frac{eB}{m^{2}}\right) ,  \label{15} \\
\frac{d{\cal W}_{P}}{dm}=- &&\frac{eB}{2\pi ^{3/2}a}\int\limits_{m^{2}/
\Lambda ^{2}}^{\infty }\frac{ds}{s^{1/2}}e^{-s}\sum\limits_{n=0}^{\infty
}e^{-s\frac{4\pi ^{2}n^{2}}{\left( am\right) ^{2}}}\coth \left( s\frac{eB} {%
m^{2}}\right) ,  \label{16}
\end{eqnarray}
where we have used again the series representation of the Jacobi functions $%
\theta _{2}$ and $\theta_{3}$. Evaluating Eqs. (\ref{15}) and (\ref{16}) at $%
m=0$ we obtain for $eB\neq 0$ 
\begin{equation}
\left\langle 0\left| \overline{\psi }\psi \right| 0\right\rangle _{AP}=0,
\label{17}
\end{equation}

\begin{equation}
\left\langle 0\left| \overline{\psi }\psi \right| 0\right\rangle _{P} =-%
\frac{eB}{2\pi a}.  \label{18}
\end{equation}
Thus, from these results it can be seen that in the {\it PBC} case a chiral
condensate is catalyzed by the magnetic field (Eq. (\ref{18})), while for
the {\it APBC }case the condensate is absent (Eq. (\ref{17})).

To understand better the genesis of the different behavior of these two
cases we should underline that the main difference between the {\it APBC}
and the {\it \ PBC }configurations lies in the existence of a zero fermion
mode $\left( n=0\right) $ for the last one. That is, for {\it PBC }the
fermion spectrum accepts a zero mode, as discussed below Eq. (\ref{7}). This
zero mode contribution is essential to obtain a different from zero result (%
\ref{18}).

We must also stress that the condensate (\ref{18}) is in fact equal to the
product of the condensate $\left( -\frac{eB}{2\pi }\right) $, found in a
topologically trivial (2+1)-dimensional domain in the presence of an
external magnetic field\cite{CSB1}$,$ times $\left( 1/a\right) $. The $1/a$
factor here is a manifestation of an additional dimensional reduction
(besides that introduced by the magnetic field\cite{CSB1}) associated to the
nontrivial topology (i.e. to the presence of the boundaries determined by
the parallel plates).

When an external magnetic field is present, but the topology is trivial, the
chiral condensate in the free fermions (3+1)-dimensional case is given by $%
\left\langle 0\left| \overline{\psi }\psi \right| 0\right\rangle \sim m\ln m$
\cite{CSB2}, and therefore the condensate is absent in the ($m\rightarrow 0$%
)-limit. From Eq.(\ref{18}), we can see that when this model is considered
within a nontrivial topology $R^{3}\times S^{1}$, the magnetic field
catalyzes the chiral symmetry breaking, even though the fermions do not
acquire a dynamical mass. Therefore, we can conclude that the magnetic
catalysis of dynamical symmetry breaking in this case is specifically
settled by the nontrivial topology.

To generate a fermion dynamical mass we must introduce fermion interactions.
As shown below, in the {\it PBC }case a fermion dynamical mass, depending on
the magnetic field $B$ and the compactified dimension length $a$, appears
for weak fermion interactions.

Let us consider the Nambu-Jona-Lasinio (NJL) model\cite{NJL} 
\begin{equation}
{\cal L}_{NJL}=\frac{1}{2}\left[ \overline{\psi },i\gamma ^{\mu }D_{\mu
}\psi \right] +\frac{G}{2N}\left[ \left( \overline{\psi }\psi \right)
^{2}+\left( \overline{\psi }i\gamma ^{5}\psi \right) ^{2}\right] ,
\label{19}
\end{equation}
where $D_{\mu }$ is the covariant derivative introduced in Eq.(\ref{1}) and
the fermions carry an additional, ''color'', index $\alpha =1,2,\dots ,N$.
It is known that introducing the composite fields 
\begin{equation}
\sigma =-\frac{G}{N}\left( \overline{\psi }\psi \right) ,\qquad \pi =-\frac{G%
}{N}\left( \overline{\psi }i\gamma ^{5}\psi \right)  \label{20}
\end{equation}
the NJL Lagrangian density (\ref{19}) can be rewritten as 
\begin{equation}
{\cal L}_{NJL}=\frac{1}{2}\left[ \overline{\psi },i\gamma ^{\mu }D_{\mu
}\psi \right] -\overline{\psi }\left( \sigma +i\gamma ^{5}\pi \right) \psi -%
\frac{N}{2G}\left( \sigma ^{2}+\pi ^{2}\right) .  \label{21}
\end{equation}
Eqs.(\ref{20}) are the Euler-Lagrange equations for the auxiliary fields $%
\sigma $ and $\pi $ in (\ref{21}). Thus, with the constraints (\ref{20}) the
theories represented by the Lagrangian densities (\ref{19}) and (\ref{21})
are equivalent.

It is clear from Eq.(\ref{21}) that if $\sigma $ gets a different from zero
vacuum expectation value (vev) $\bar{\sigma}$, the fermions acquire mass.
Then, to investigate the generation of a fermion dynamical mass we need to
search for possible non-zero vev of $\sigma $. In the large $N$ limit the
vacuum is determined by the stationary point of the effective action for the
composite fields $\sigma $ and $\pi $, obtained by integrating over fermions
in the path integral: 
\begin{equation}
W(\sigma ,\pi )=-\frac{N}{2G}\int d^{4}x(\sigma ^{2}+\pi ^{2})-i{\rm Tr}\log
\left[ i\gamma ^{\mu }D_{\mu }-\left( \sigma +i\gamma _{5}\pi \right)
\right] .  \label{effaction}
\end{equation}
Since the vacuum should respect translational invariance, to find the vacuum
solution we need to calculate the effective action for constant auxiliary
fields. In this case, the effective action is just $W(\sigma ,\pi
)=-V(\sigma ,\pi )T{\cal A}a$, where $T{\cal A}a$ is the space-time volume
and $V$ is the effective potential. Moreover, since the effective potential $%
V$ depends only on the chiral invariant $\rho ^{2}=\sigma ^{2}+\pi ^{2},$ it
is sufficient to consider a configuration with $\pi =0$ and $\sigma $
constant. In the proper-time formalism we get

\begin{eqnarray}
V_{AP}(\sigma ) &=&\frac{N\sigma ^{2}}{2G}+\frac{NeB}{8\pi ^{2}}%
\int\limits_{1/\Lambda ^{2}}^{\infty }\frac{ds}{s^{2}}e^{-s\sigma
^{2}}\theta _{4}\left( 0\left| \frac{ia^{2}}{4\pi s}\right. \right) \coth
(eBs),  \label{VAP} \\
V_{P}(\sigma ) &=&\frac{N\sigma ^{2}}{2G}+\frac{NeB}{8\pi ^{2}}%
\int\limits_{1/\Lambda ^{2}}^{\infty }\frac{ds}{s^{2}}e^{-s\sigma
^{2}}\theta _{3}\left( 0\left| \frac{ia^{2}}{4\pi s}\right. \right) \coth
(eBs),  \label{VP}
\end{eqnarray}
for antiperiodic and periodic boundary conditions respectively. It is
evident that in the limit $a\rightarrow \infty $, since the theta functions $%
\theta _{3,4}(0|ia^{2}/{4\pi s})|_{a=\infty }=1$, the two effective
potentials are equal $(V_{AP}=V_{P})$ and coincide with the well known
effective potential of the NJL model in the presence of an external magnetic
field \cite{CSB2}.

The dynamical mass $\bar{\sigma}$ for the {\it PBC} case is obtained as the
solution of the gap equation $dV_{P}/{d\sigma }=0$. This equation can be
written separating the ultraviolet contribution (i.e. terms depending on $%
\Lambda $). In leading order in $1/\Lambda $ the PBC gap equation is given
by 
\begin{eqnarray}
&&\sigma \left[ {\frac{1}{G}}-\frac{\Lambda ^{2}}{4\pi ^{2}}+\frac{\sigma
^{2}}{4\pi ^{2}}\left( \log \frac{\Lambda ^{2}}{\sigma ^{2}}+1-\gamma
\right) -\frac{1}{4\pi ^{2}}\int\limits_{0}^{\infty }\frac{ds}{s^{2}}%
e^{-s\sigma ^{2}}\left( \theta _{3}\left( 0\left| \frac{ia^{2}}{4\pi s}%
\right. \right) -1\right) \right.   \nonumber \\
&&\left. -\frac{eB}{4\pi ^{2}}\int\limits_{0}^{\infty }\frac{ds}{s}%
e^{-s\sigma ^{2}}\theta _{3}\left( 0\left| \frac{ia^{2}}{4\pi s}\right.
\right) \left( \coth (eBs)-{\frac{1}{eBs}}\right) +O\left( \frac{1}{\Lambda }%
\right) \right] =0,  \label{gapequation}
\end{eqnarray}
where $\gamma \approx 0.577$ is the Euler constant. The corresponding gap
equation for {\it APBC} is obtained by replacing $\theta _{3}$ by $\theta
_{4}$ in Eq. (\ref{gapequation}).

It is easy to see that under the condition $1/a\ll \sigma \ll \sqrt{eB}$,
i.e. $a$ being the largest length scale in the problem, the gap equations
for both cases ({\it PBC }and {\it APBC}) reduce to the following one 
\begin{equation}
\sigma \left[ \frac{1}{G}-\frac{1}{G_{c}}\pm \left( \frac{2\sigma }{\pi a^{3}%
}\right) ^{1/2}e^{-\sigma a}-\frac{eB}{4\pi ^{2}}\log \frac{eB}{\pi \sigma
^{2}}\right] =0,  \label{AA}
\end{equation}
where $G_{c}=({4\pi ^{2}}/\Lambda ^{2})$ and $\pm $ refers to {\it APBC} and 
{\it PBC}, respectively. The solution of Eq. (\ref{AA}) in the $G\ll G_{c}$
approximation is 
\begin{equation}
m_{dyn}\equiv \bar{\sigma}=\sqrt{\frac{eB}{\pi }}\exp \left( -\frac{2\pi ^{2}%
}{eBG}\right) .  \label{PLmass}
\end{equation}
As expected, the solution (\ref{PLmass}), which is nonanalytic in $G$ as $%
G\to 0,$ coincides with the one found in (3+1)-dimensions for $B\neq 0$ and $%
a=\infty $ \cite{GMSPL}.

Let us consider now the opposite limit of the small length $a$ ($\sigma, 
\sqrt{eB} \ll 1/a$) which is the important one to study the effects of the
compactified dimension. In this limit the gap equation (\ref{gapequation})
for {\it PBC} reduces to 
\begin{equation}
\sigma \left[ {\frac{1}{G}}-\frac{1}{G_{c}}-\frac{1}{3a^{2}}+\frac{\sigma
^{2}}{4\pi ^{2}}\left( \log \frac{\Lambda ^2a^2}{16\pi^2}+\gamma \right) -%
\frac{eB}{2\pi \sigma a}\left( \sqrt{\frac{2}{eB}}\sigma \zeta ({\frac{1}{2}}%
,\frac{\sigma ^{2}}{2eB}+1)+1\right) \right] =0,  \label{gapeq:PBC}
\end{equation}
where $\zeta (\nu ,x)$ is the generalized Riemann zeta function.

As $B\rightarrow0$, we recover the known gap equation \cite{Klimenko} which
admits a nontrivial solution only if the coupling is supercritical, $G>G_c^a$%
, and the critical coupling $G_{c}^{a}=( G_{c}^{-1}+1/3a^{2})^{-1}$. When an
external magnetic field, $B\neq0$, is present a nontrivial solution exists
at all $G>0$ and, in particular, at $G\ll G_c^a$. Indeed, looking at the
solution of Eq. (\ref{gapeq:PBC}) satisfying $\bar{\sigma}\ll \sqrt{eB}$ we
find 
\begin{equation}
m_{dyn}^{{\it P}}\equiv \bar{\sigma}\simeq \frac{eB}{2\pi a} \frac{GG_{c}^{a}%
}{G_{c}^{a}-G}  \label{38}
\end{equation}
if the coupling $G\ll G_{c}^{a}$. The condition $G<G_{c}^{a}$ guarantees
that (\ref{38}) is a minimum solution of the effective potential $V_{P}$.

From the above result it is clear that the dynamical mass solution (\ref{38}%
) exists in the weak coupling regime of the theory. The fact that there is
no critical value of the coupling to produce chiral symmetry breaking is a
characteristic feature of the catalysis of dynamical symmetry breaking by a
magnetic field\cite{CSB1,GMSPL,CSB2}. It is remarkable that unlike the $%
a=\infty $ case, where the dynamical mass has nonanalytical dependence on
the coupling constant as $G\to 0$ (see Ref.\cite{GMSPL}), at finite $a$ the
dynamical mass (\ref{38}) is an analytic function of $G$ at $G=0$.

Note also the following point. Equation (\ref{20}) implies that $m_{{\rm dyn}%
}=\langle 0|\sigma |0\rangle =-G\langle 0|\bar{\psi}\psi |0\rangle /N$. From
here and Eq. (\ref{38}) we find that the condensate $\langle 0|\bar{\psi}%
\psi |0\rangle $ is $\langle 0|\bar{\psi}\psi |0\rangle =-N|eB|/{2\pi a}$ in
leading order in $G$; i.e. it coincides with the value of the condensate
calculated in the problem of free fermions in a magnetic field (see Eq. (\ref
{18})). This point also explains why the dynamical mass $m_{{\rm dyn}}$ is
an analytic function of $G$ at $G=0$: indeed, the condensate already exists
at $G=0$. As a result, we have big enhancement of the dynamical fermion mass
generation in the presence of boundaries along the magnetic field direction
and with periodic boundary conditions for the fermion fields comparing to
the case of topologically trivial space-time (see Eqs. (\ref{38}) and (\ref
{PLmass})).

As it is well known, the breaking of a continuous chiral symmetry is linked
to the existence of Nambu-Goldstone (NG) bosons. The analysis of the
NG-modes appearing in this problem will be published elsewhere.

Let us discuss now the {\it APBC} case. Following the same procedure we used
for {\it PBC}, it is easy to show that the {\it APBC} gap equation under
conditions $\sigma ,\sqrt{eB}\ll 1/a$ does not have a nontrivial solution;
on the other hand, when $\sigma \ll \sqrt{eB},1/a$ it is reduced to 
\begin{eqnarray}
&&\sigma \left[ {\frac{1}{G}}-\frac{1}{G_{c}}+\frac{1}{6a^{2}}-\frac{eB}{4
\pi^{2}}\int\limits_{0}^{\infty }\frac{ds}{s}\theta_{4}\left( 0\left| \frac{i%
}{4\pi s}\right. \right) \left(\coth (eBa^{2}s) -{\frac{1}{eBa^{2}s}}%
\right)\right.  \nonumber \\
&&+\left.\frac{\sigma^{2}}{4\pi ^{2}}\left( \log \frac{\Lambda ^{2}a^2}{%
\pi^{2}}+\gamma +eBa^2\int\limits_{0}^{\infty }ds\theta_{4}\left( 0 \left| 
\frac{i}{4\pi s}\right. \right) \left( \coth (eBa^{2}s) -{\frac{1}{eBa^{2}s}}%
\right)\right) \right] =0.  \label{gapeq:APBC}
\end{eqnarray}
From Eq. (\ref{gapeq:APBC}) one can convince oneself that there is no
nontrivial solution under the assumptions made if the coupling is weak ($%
G\to 0$). For chiral symmetry breaking to take place, the coupling constant $%
G$ must be larger than some critical value that depends on the magnitude of
the magnetic field $B$ and size $a$. Indeed, Eq.(\ref{gapeq:APBC}) can be
simplified in the limiting case $\sqrt{eB}a\gg 1$ 
\begin{equation}
\sigma \left[ \frac{1}{G}-\frac{1}{G_{c}}+\frac{1}{6a^{2}}-\frac{eB}{4\pi
^{2}}\left( \ln \frac{eBa^{2}}{\pi ^{3}}+2\gamma \right) + \frac{\sigma ^{2}%
}{4\pi ^{2}}\left( \log \frac{\Lambda ^{2}a^2}{\pi^{2}} +\frac{7\zeta(3)}{%
4\pi^2}eBa^2+\gamma \right) \right] =0.  \label{eq:lla}
\end{equation}
From Eq. (\ref{eq:lla}) one can notice that the magnetic field is helping
the symmetry breaking since the critical coupling is less than the one
corresponding to the case with zero magnetic field.

On the other hand, the contribution of the compactified dimension length $a$
to the gap equation in the presence of a magnetic field has opposite sign
for {\it APBC} (third term in Eq. (\ref{gapeq:APBC})), as compared to {\it %
PBC} (third term in Eq. (\ref{gapeq:PBC})). Consequently, the boundary
effect in the {\it APBC} case is not enhancing the chiral symmetry breaking,
but on the contrary, it is counteracting it; while in the {\it PBC} case the
magnetic catalysis is substantially enhanced by the boundary.

The fact that the boundary effect in the {\it APBC} case is not enhancing
the chiral symmetry breaking can be better understood if one realizes that
due to the antiperiodic boundary conditions the quantity $1/a$ plays in the 
{\it APBC} case a role similar to temperature. We have seen above that the
chiral symmetry breaking takes place at small $1/a$ with a corresponding
dynamical mass given by Eq. (\ref{PLmass}), so we should expect that at $1/a$
larger than some critical value $1/a_{c}$ the chiral symmetry must be
restored.

Such a critical value $1/a_{c}$ indeed exists and is determined from the
condition that the second derivative of the effective potential at $\sigma=0$
becomes positive. In fact, from Eq. (\ref{eq:lla}) it is found to be 
\begin{equation}
\frac{1}{a_{c}}=\frac{e^{\gamma }}{\pi }m_{dyn}  \label{ac}
\end{equation}
with $m_{dyn}$ the dynamical mass (\ref{PLmass}). Therefore, we obtain that
the inverse of the critical length, $1/a_{c}$, is of the order of $m_{dyn}$,
a result equivalent to that found for the relationship between the critical
temperature and the gap in BCS superconductivity. Also, it follows from Eq. (%
\ref{eq:lla}) that the approach of $\bar\sigma=m_{dyn}$ to zero is given by
the square root dependence $(a-a_c)^{1/2}$.

Finally, we should point out that the combined effect of an external
magnetic field and periodic boundary conditions for fermions along third
axis can find important applications to explaining a kink-like feature of
thermal conductivity in the presence of a magnetic field in high-$T_{c}$
superconducting samples much below the critical temperature. Such an effect
was observed in recent experiments by Krishana et al. \cite{Krishana}. These
high-$T_{c}$ superconductors are known to possess a quasi-2D structure, and
it has been suggested \cite{Krishana,CM} that the magnetic field can induce
a second phase transition in the superconducting state, leading to the
opening of a gap at the nodes of a conventional $d$-wave gap.

\begin{acknowledgments}
We are grateful to V. A. Miransky and I. A. Shovkovy for useful
remarks. This research has been supported in part by the National
Science Foundation under Grant No. PHY-9722059.
\end{acknowledgments}

\end{document}